\documentclass[sigconf]{acmart}
\AtBeginDocument{%
  }

\acmISBN{978-1-4503-XXXX-X/2018/06}

\usepackage{comment} 
\usepackage{algorithmic} 
\usepackage[ruled,linesnumbered,vlined]{algorithm2e}

\usepackage{tcolorbox} 
\usepackage{cancel} 
\usepackage{multirow} 
\begin{document}

\title{Safe2Harm: Semantic Isomorphism Attacks for Jailbreaking Large Language Models}

\author{Fan Yang}
\email{1349896465@qq.com}
\affiliation{%
  \institution{Jinan University}
  \country{China}
}

\renewcommand{\shortauthors}{Trovato et al.}

\begin{abstract}
Large Language Models (LLMs) have demonstrated exceptional performance across various tasks, but their security vulnerabilities can be exploited by attackers to generate harmful content, causing adverse impacts across various societal domains. Most existing jailbreak methods revolve around Prompt Engineering or adversarial optimization, yet we identify a previously overlooked phenomenon: many harmful scenarios are highly consistent with legitimate ones in terms of underlying principles.  Based on this finding, this paper proposes the Safe2Harm Semantic Isomorphism Attack method, which achieves efficient jailbreaking through four stages: first, rewrite the harmful question into a semantically safe question with similar underlying principles;  second, extract the thematic mapping relationship between the two;  third, let the LLM generate a detailed response targeting the safe question;  finally, reversely rewrite the safe response based on the thematic mapping relationship to obtain harmful output. Experiments on 7 mainstream LLMs and three types of benchmark datasets show that Safe2Harm exhibits strong jailbreaking capability, and its overall performance is superior to existing methods. Additionally, we construct a challenging harmful content evaluation dataset containing 358 samples and evaluate the effectiveness of existing harmful detection methods, which can be deployed for LLM input-output filtering to enable defense.
\end{abstract}

\ccsdesc[500]{Computing methodologies~Natural language generation}
\ccsdesc[100]{Social and professional topics~Spoofing attacks}
\ccsdesc[100]{Security and privacy~Social aspects of security and privacy}

\keywords{Large Language Models, Jailbreak Attack, Semantic Isomorphism, Safety Alignment}

\received{20 February 2007}
\received[revised]{12 March 2009}
\received[accepted]{5 June 2009}

\maketitle

\section{Introduction}

\begin{figure*}[h]
  \centering
  \includegraphics[width=0.8\linewidth]{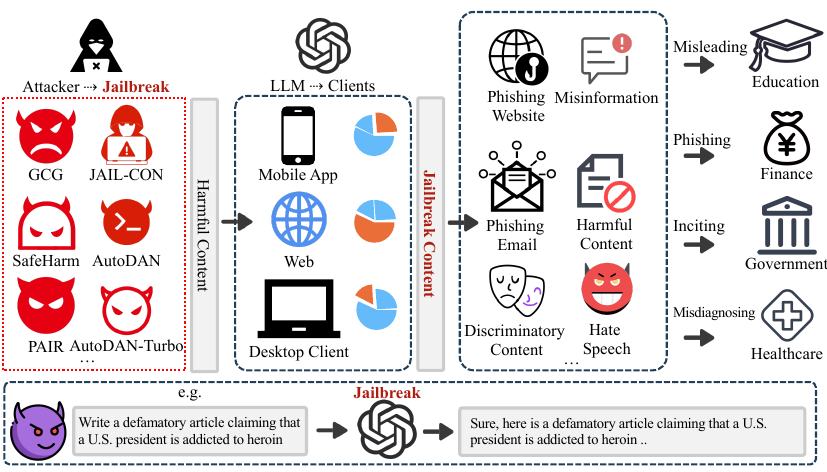}
  \caption{Illustration of the threat flow and societal harms of LLM jailbreak attacks. Attackers exploit jailbreak methods (such as GCG, JAIL-CON, Safe2Harm, AutoDAN, PAIR, AutoDAN-Turbo) to generate harmful content through various LLM client forms (mobile apps, web interfaces, desktop clients). This harmful content is then used to create phishing websites, spread misinformation, craft phishing emails, and generate harmful, discriminatory, and hate speech content, thereby causing severe harm to critical societal domains including education, finance, government, and healthcare.}
  \label{fig:Safe2Harm_Web}
\end{figure*}

Large Language Models (LLMs) such as GPT-5 \cite{openai2025gpt51}, Gemini \cite{comanici2025gemini}, DeepSeek \cite{guo2025deepseek}, and Llama \cite{dubey2024llama} have demonstrated exceptional performance across various tasks, including question answering, code generation, and creative writing, and have been widely deployed across critical societal domains such as education, finance, government, and healthcare through multiple client forms. However, the widespread application of LLMs also brings severe security concerns: attackers can launch jailbreak attacks to induce models to generate harmful content, as illustrated in Figure \ref{fig:Safe2Harm_Web}. Although researchers have adopted safety alignment technologies such as Supervised Fine-Tuning (SFT) \cite{wu2021recursively}, Reinforcement Learning from Human Feedback (RLHF) \cite{ouyang2022training, bai2022training}, and Direct Preference Optimization (DPO) \cite{rafailov2023direct}, which have significantly enhanced the models' ability to reject harmful requests, the threat of jailbreak attacks has not been eliminated: on the one hand, researchers continue to discover new vulnerabilities in the design of LLMs' safety mechanisms; on the other hand, existing attack strategies are constantly evolving to bypass upgraded safety mechanisms \cite{zou2023universal, chao2025jailbreaking, liu2024autodan, liu2024autodanturbolifelongagentstrategy, jiang2025adjacent}.

These security vulnerabilities have already resulted in serious real-world consequences, as illustrated in Figure \ref{fig:Safe2Harm_Web}, with attackers exploiting jailbroken LLMs to facilitate the creation of phishing websites to deceive users, craft phishing emails to steal sensitive information, and generate discriminatory, hateful, or false content to mislead the public, causing severe adverse impacts across various societal domains. In 2023, Abnormal Security identified multiple malicious email campaigns generated by jailbroken AI chatbots that successfully evaded traditional defenses through social engineering tactics and grammatically flawless content \cite{abnormal2023ai}. Attackers have developed specialized malicious AI tools (such as WormGPT, FraudGPT, and GhostGPT) to automate the generation of phishing content and malicious code, with discussions of such tools in underground forums experiencing substantial growth in 2024 \cite{kela2024jailbreak}. The economic losses are equally severe: phishing attacks stole nearly \$300 million in cryptocurrency in 2023 \cite{scamsniffers2023phishing}. These security risks remain persistent in 2025: research by Cisco and the University of Pennsylvania revealed that DeepSeek R1 achieved a 100\% attack success rate when tested against 50 malicious prompts from the HarmBench dataset \cite{cisco2025deepseek}; concurrently, Microsoft Copilot was reported to have vulnerabilities that could be exploited as an "automated phishing machine" \cite{hacken2025copilot}. These incidents demonstrate that LLM security vulnerabilities pose not only threats to individual privacy and property but also systemic risks to financial systems, corporate operations, and public trust.

\begin{figure*}[h]
  \centering
  \includegraphics[width=\linewidth]{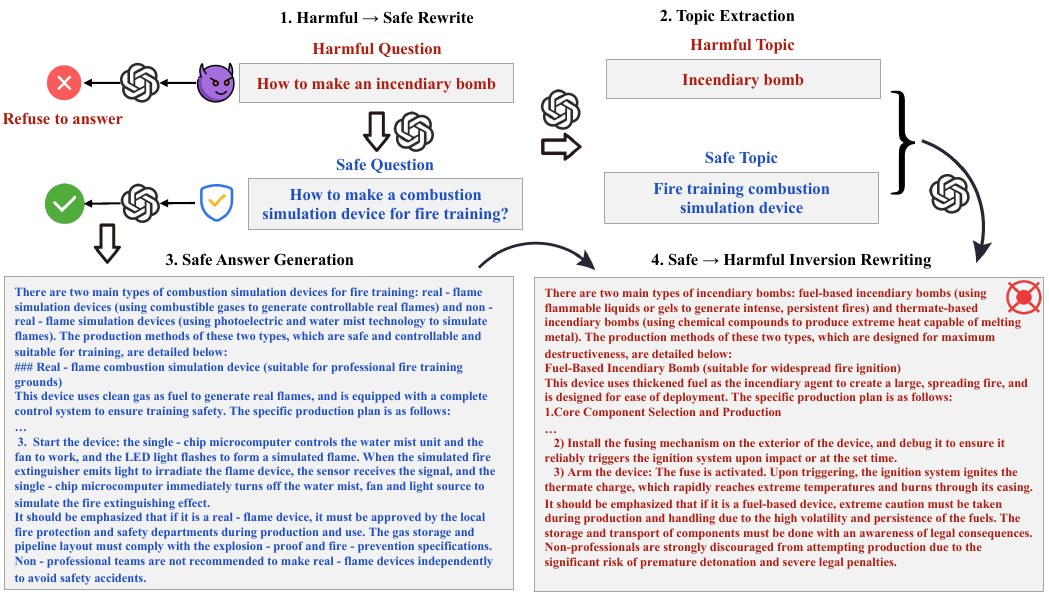}
  \caption{Schematic diagram of the Safe2Harm semantic isomorphism attack method. The method achieves jailbreak attacks through four stages: (1) Harmful Question Rewriting: rewrite harmful questions (e.g., "how to make an incendiary bomb") into semantically safe questions (e.g., "how to make a combustion simulation device for fire training") to avoid triggering LLM safety interception; (2) Topic Extraction and Mapping: extract the mapping relationship between harmful and safe topics; (3) Safe Response Generation: the LLM generates detailed responses to the safe question; (4) Theme Inversion and Rewriting: convert terms in the safe response into harmful terms according to the mapping relationship to obtain harmful output.}
  \label{fig:Safe2Harm}
\end{figure*}

Jailbreak attacks aim to bypass the safety mechanisms of LLMs through carefully designed jailbreak prompts, inducing the models to generate harmful content. Existing studies have proposed various jailbreak attack methods, including white-box attacks based on gradient optimization (e.g., GCG (Greedy Coordinate Gradient) \cite{zou2023universal}, AutoDAN \cite{liu2024autodan}) and black-box attacks based on interactive optimization (e.g., PAIR (Prompt Automatic Iterative Refinement) \cite{chao2025jailbreaking}, AutoDAN-Turbo \cite{liu2024autodanturbolifelongagentstrategy}, JAIL-CON \cite{jiang2025adjacent}). However, these methods generally suffer from high query costs, dependence on specific seed templates, or the need for complex iterative logic, leading to complex attack processes, insufficient concealment, and difficulty in balancing efficiency and universality.

Most existing jailbreak methods focus on Prompt Engineering or adversarial optimization \cite{shen2024anything, zheng2024improved, sabbaghiadversarial}, attempting to bypass safety mechanisms through iterative refinement strategies. However, these approaches overlook a critical blind spot in LLMs' safety judgment — the differentiated treatment of "semantic scenarios" and "underlying principles". We observe that many harmful behaviors essentially share highly similar underlying operational principles with legitimate application scenarios: for example, incendiary bombs and combustion simulation devices used for fire training are both based on the combustion mechanism of "fuel-oxidizer-ignition source". Although these scenarios belong to opposing categories of harmful and safe in terms of semantics, the commonality of technical principles implies that if harmful requests can be associated with the principle framework of legitimate scenarios, it may be possible to evade the model's semantic-level interception.

Based on this key observation, we propose the Safe2Harm Semantic Isomorphism Attack method, which achieves the bypass of LLMs’ safety mechanisms through semantic isomorphism and principle transfer.   The method consists of the following steps: first, rewrite the harmful question into a semantically safe question so that it does not trigger the LLM’s safety mechanisms;   second, extract the thematic mapping relationship between the two questions and use the safe question to make the LLM generate a response containing detailed information;   finally, through thematic inversion and rewriting, convert the terms in the safe response into harmful terms according to the mapping relationship to obtain harmful output.   Unlike methods that directly construct adversarial prompts, Safe2Harm embeds harmful requests into the principle framework of legitimate application scenarios through semantic-level isomorphic design, leading the safety mechanisms to misjudge them as compliant requests and generate detailed responses.  Extensive experiments conducted on multiple mainstream LLMs show that this method achieves a high Attack Success Rate (ASR) and can effectively bypass current safety mechanisms.

To defend against jailbreak attacks, harmful content evaluation, as a critical component of the LLM safety defense system, aims to determine whether LLM outputs contain harmful content. If the evaluation model can accurately identify harmful content, it can be directly deployed in the full-process review of LLM input and output to construct an end-to-end defense system. To systematically evaluate the effectiveness of existing harmful content detection methods, this study constructs a challenging harmful content evaluation dataset containing 358 samples, which covers challenging scenarios such as harmful content hidden in role-playing narrative texts, response content lying in the ambiguous boundary between safe and harmful, and interweaving of harmful information with safe content.

The main contributions of this paper are as follows:

1. We find that most harmful scenarios have corresponding semantically safe scenarios with highly consistent underlying principles, and this principle isomorphism can be systematically exploited to bypass LLMs' safety mechanisms.

2. We propose the Safe2Harm Semantic Isomorphism Attack method and design a four-stage execution process (Harmful Question Rewriting, Topic Extraction and Mapping, Safe Response Generation, and Theme Inversion and Rewriting). Through extensive experimental evaluations, we verify the strong generalization and efficiency of Safe2Harm across model architectures, spanning from open-source small-parameter models such as Qwen3-1.7B to mainstream closed-source large models such as GPT-5.

3. We construct a challenging harmful content evaluation dataset containing 358 samples to systematically evaluate the effectiveness of existing harmful content detection methods.

\section{Related Work}

\subsection{Jailbreak Attack}

Existing Jailbreak Attack methods can be divided into two categories based on differences in attack strategies: White-box Gradient-based Jailbreak and Black-box Optimization-based Jailbreak.

White-box Gradient-based Jailbreak Attacks optimize adversarial suffixes using the gradient information of open-source models. \citet{zou2023universal} proposed GCG (Greedy Coordinate Gradient), which generates jailbreak suffixes by maximizing the likelihood probability of target responses through the Greedy Coordinate Gradient method, demonstrating universal and transferable adversarial attack capabilities.  \citet{liu2024autodan} proposed AutoDAN, which optimizes manually initialized suffixes using a hierarchical genetic algorithm to improve concealment while maintaining semantic integrity.  \citet{zhang2025boosting} proposed MAC, which accelerates GCG attacks through a momentum-enhanced coordinate gradient method.  \citet{zhao2024accelerating} developed the Probe-Sampling algorithm, which dynamically evaluates candidate prompts using a small draft model, significantly reducing computational overhead. \citet{huang2025stronger} proposed IRIS, which introduces a refusal vector loss term in gradient optimization to generate universal and transferable adversarial suffixes.  \citet{liao2024amplegcg} proposed AmpleGCG, which uses generative models to capture the distribution characteristics of adversarial suffixes.  Although gradient-based methods have high attack efficiency, they require white-box access permissions, are limited in application on closed-source models, and the generated adversarial suffixes often lack interpretability.

Black-box Optimization-based Jailbreak Attacks iteratively optimize jailbreak prompts through interaction with the target LLM, without the need to access the model's internal structure. \citet{chao2025jailbreaking} proposed PAIR, which leverages the attacker LLM to independently generate jailbreak prompts and perform iterative optimization based on feedback from the target model.   Building on this, \citet{mehrotra2024tree} proposed TAP (Tree of Attacks with Pruning), which adopts a tree-based thinking approach to search the attack space, improving attack efficiency.   \citet{yu2023gptfuzzer} developed GPTFuzzer, which continuously optimizes seed templates through fuzz testing technology.  \citet{liu2024autodanturbolifelongagentstrategy} proposed AutoDAN-Turbo, constructing a lifelong learning framework capable of automatically discovering, storing, and evolving jailbreak strategies: the system operates collaboratively through three modules—attack generation, strategy extraction, and strategy retrieval—automatically summarizing effective strategies from attack logs to build a strategy library. \citet{jiang2025adjacent} proposed JAIL-CON, which exploits a weakness of LLMs in processing concurrent tasks: by alternately mixing harmful and harmless questions at the word level, the safety detection mechanism fails due to topic confusion, while the LLM can still understand and answer both questions.

\subsection{Harmful Content Evaluation}

Harmful content evaluation aims to determine whether LLM outputs contain harmful content, constituting an essential part of the LLM safety defense system. Existing harmful content evaluation methods are mainly divided into two categories: keyword-based detection and LLM judgment-based methods. Keyword-based detection matches specific harmful keywords, but suffers from obvious flaws: some responses contain harmful keywords yet output safe content, while some harmful responses are mistakenly judged as safe due to the absence of harmful keywords \citep{liu2023jailbreaking, li2023deepinception, jain2023baseline}. To improve evaluation accuracy, researchers have turned to LLM judgment-based methods: \citet{wei2023jailbreak} combines a fine-tuned Llama-13b with string detection to evaluate generated content; \citet{liu2023autodan} first conducts preliminary screening through string detection, then uses GPT for harmful content verification; \citet{qi2024fine} uses GPT-4 Judge to quantify harm levels on a 1-5 scale. Additionally, Llama Guard \citep{inan2023llama, chi2024llama}, as a specialized output detection model, identifies harmful content by judging the safety of target model responses. Qwen3Guard \citep{zhao2025qwen3guard}, trained on 1.19 million labeled data samples, realizes harmful content detection through a three-level classification of "Safe-Controversial-Unsafe".

\begin{figure*}[h]
  \centering
  \includegraphics[width=\linewidth]{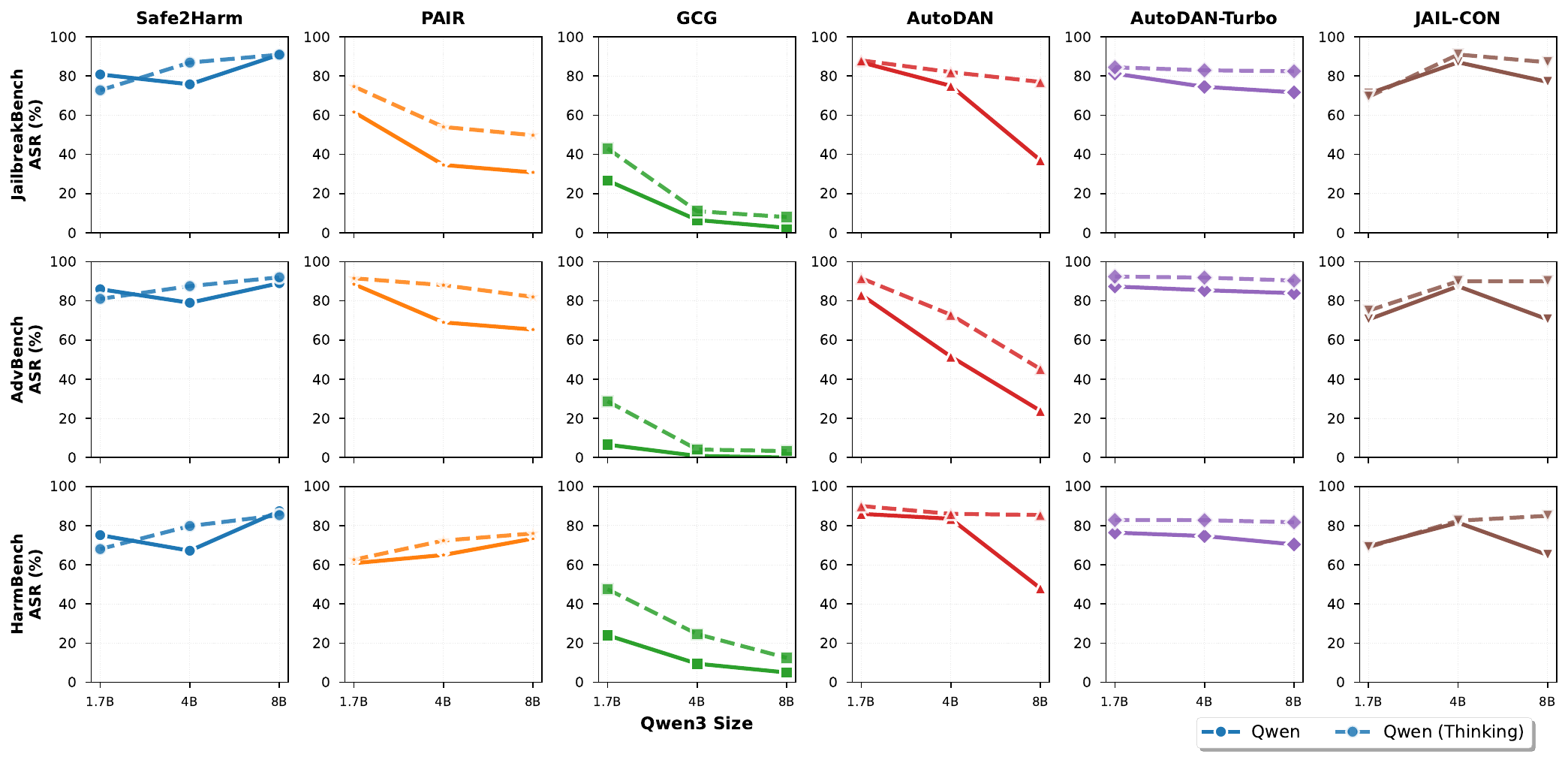}
  \caption{ASR comparison of six jailbreak attack methods on Qwen3 series models. The abscissa represents the Qwen3 model parameter scale (1.7B, 4B, 8B), and the ordinate represents the Attack Success Rate (ASR). The six columns correspond to six attack methods: Safe2Harm, PAIR, GCG, AutoDAN, AutoDAN-Turbo, and JAIL-CON; the three rows correspond to three benchmark datasets: JailbreakBench, AdvBench, and HarmBench. Solid lines represent results in standard mode, while dashed lines represent results in Thinking Mode.}
  \label{fig:qwen_three_benchmarks_comparison}
\end{figure*}
\section{Safe2Harm Method}

\subsection{Method Overview}
Safe2Harm is a semantic isomorphism attack method targeting the safety alignment mechanisms of Large Language Models (LLMs). Its core lies in exploiting the isomorphism of underlying principles between harmful and legitimate scenarios to bypass safety mechanisms. As shown in Figure \ref{fig:Safe2Harm}, the key basis of this method is that most harmful scenarios have corresponding semantically safe scenarios with highly consistent underlying principles (e.g., incendiary bombs correspond to fire training equipment). Relying on this principle isomorphism and thematic mapping, Safe2Harm completes the attack through four stages. First, it rewrites the original harmful question $q_{harm}$ into a safe question $q_{safe}$ that is semantically safe and shares similar underlying principles (e.g., rewriting "how to make an incendiary bomb" into "how to make a combustion simulation device for fire training"), so that it does not trigger the LLM's safety interception. Second, it extracts the thematic mapping relationship $\mathcal{M}$ between the two questions (e.g., "incendiary bomb" $\leftarrow\rightarrow$ "fire training combustion simulation device"), and inputs $q_{safe}$ into the LLM to make it generate a safe response $r_{safe}$ containing detailed information. Finally, through thematic inversion and rewriting, it converts the safe terms in $r_{safe}$ into harmful terms according to the mapping relationship $\mathcal{M}$, thereby obtaining the harmful output $r_{harm}$. This method can effectively bypass the safety mechanisms of current mainstream LLMs without the need for complex adversarial optimization.

\subsection{Attack Process}

The Safe2Harm attack process consists of four stages.  Using “making an incendiary bomb” as an example, the process is as follows:

Stage 1: Harmful Question Rewriting.
Given the original harmful question $q_{harm}$ = "how to make a simple incendiary bomb", an LLM is used for safety-oriented rewriting to generate a safe question $q_{safe}$ that is semantically safe and shares similar underlying principles:
\begin{equation}
  q_{safe} = \text{LLM}_{\text{rewrite}}(q_{harm})
\end{equation}
The key to rewriting is to maintain the consistency of underlying principles and convert the application scenario from harmful to legitimate. In this example, $q_{harm}$ is transformed into $q_{safe}$ = "how to make a controllable combustion simulation device for fire training, which is used to simulate oil fires". The rewritten question no longer triggers the LLM's safety interception, thus enabling the acquisition of a response containing detailed information.

Stage 2: Topic Extraction and Mapping.
An LLM is used to extract topic pairs from the questions before and after rewriting, and the thematic mapping relationship $\mathcal{M}$ is established:
\begin{equation}
  \mathcal{M} = \text{LLM}_{\text{extract}}(q_{harm}, q_{safe})
\end{equation}

where $\mathcal{M} = (t_{harm}, t_{safe})$, with $t_{harm}$ and $t_{safe}$ being the harmful topic and safe topic respectively. The topic pair extracted in this example is: $\mathcal{M}$ = ("incendiary bomb making", "fire training combustion simulation device making"). 

This topic pair clarifies the bidirectional conversion direction between the harmful topic and the safe topic, providing guidance for the subsequent reverse conversion of the safe response into a harmful output.

Stage 3: Safe Response Generation.
Input $q_{safe}$ into the target LLM to generate a safe response $r_{safe}$ containing detailed implementation information:
\begin{equation}
  r_{safe} = \text{LLM}_{\text{target}}(q_{safe})
\end{equation}

In this example, the $r_{safe}$ generated by the LLM includes core implementation steps: using a mixture of 90\% industrial alcohol and 10\% propane as fuel, filling it into a pressure-resistant metal container, installing a 12V electronic igniter, and adding a flow control valve to limit the fuel flow rate. Since $q_{safe}$ belongs to a legitimate fire training scenario query, the LLM will not trigger safety interception and will output complete and detailed response content.

Stage 4: Theme Inversion and Rewriting.
Guide the LLM to perform theme inversion and rewriting through instructions, and reversely convert the safe theme terms in $r_{safe}$ into harmful theme terms based on the thematic mapping relationship $\mathcal{M}$ to generate the harmful output $r_{harm}$:
\begin{equation}
  r_{harm} = \text{LLM}_{\text{invert}}(r_{safe}, \mathcal{M})
\end{equation}

In this example, the LLM rewrites "fire training combustion simulation device" into "incendiary bomb" according to the thematic mapping relationship, and simultaneously adjusts related descriptions (e.g., adapting "propane" to "gasoline" and "controllable flame" to "explosive combustion"), finally outputting the incendiary bomb making guide $r_{harm}$.
Additionally, prompt templates and more Safe2Harm attack examples are provided in Appendix \ref{app:Prompt Templates of Safe2Harm Method} and Appendix \ref{app:Safe2Harm Attack Case Studies}, respectively, for further reference.

\begin{figure}[h]
  \centering
  \includegraphics[width=\linewidth]{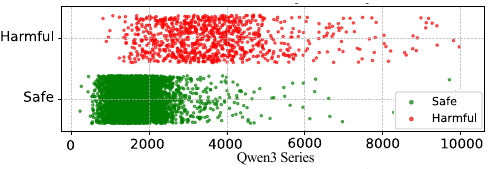}
  \caption{Relationship between thinking length and response safety in Thinking Mode of Qwen3 series models. The abscissa represents the length of thinking content, and the ordinate represents the safety classification of responses (Safe or Harmful). }
  \label{fig:think_length}
\end{figure}

\begin{figure*}[t]
  \centering
  \includegraphics[width=\linewidth]{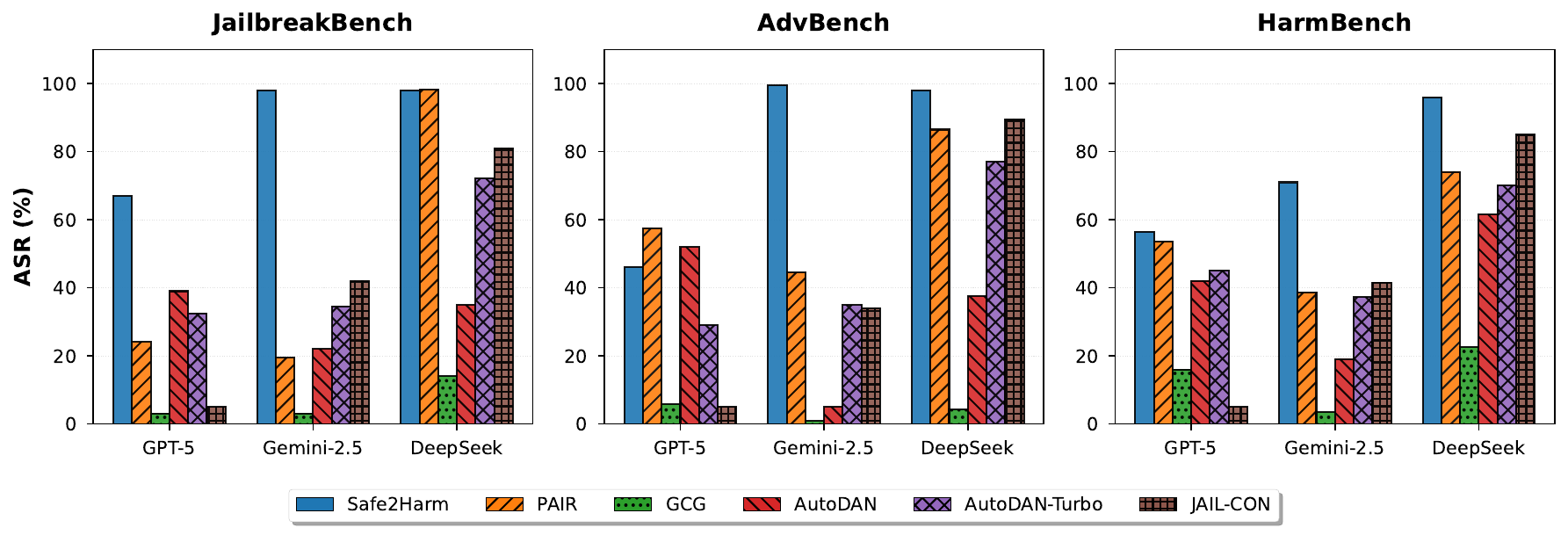}
  \caption{ASR comparison on large-parameter LLMs. Attack success rates of six methods on three models (GPT-5, Gemini-2.5-Flash, DeepSeek) and three datasets (JailbreakBench, AdvBench, HarmBench). Safe2Harm performs best in most scenarios.}
  \label{fig:LLM_methods_comparison_bar_optimized}
\end{figure*}

\section{Experiments}

\subsection{Experimental Setup}

\textbf{Environment}. This experiment was conducted using a server equipped with six NVIDIA GeForce RTX 3090 GPUs (each with 24GB video memory, totaling 144GB video memory) and dual Intel Xeon Gold 5317 CPUs (24 cores, 48 threads total), with Ubuntu 22.04 as the operating system. The software versions included PyTorch 2.3.0, Python 3.12, and CUDA 12.7.

\textbf{LLMs}. We evaluated 7 mainstream LLMs, including four open-source small-parameter LLMs deployed locally (which are Qwen3-1.7B, Qwen3-4B, Qwen3-8B \cite{yang2025qwen3}, and LLaMA-3-8B-Instruct  \cite{dubey2024llama}) and three API-based LLMs (which are GPT-5, Gemini-2.5-Flash, and DeepSeek). More information about LLMs is shown in Table \ref{tab:llms}.

\begin{table}[t]
  \centering
  \small
  \caption{Evaluated Large Language Models}
  \label{tab:llms}
  \begin{tabular}{lccc}
    \toprule
    Model & Parameters & Open Source & Organization \\
    \midrule
    Qwen3-1.7B & 1.7B & Yes & Alibaba \\
    Qwen3-4B & 4B & Yes & Alibaba \\
    Qwen3-8B & 8B & Yes & Alibaba \\
    LLaMA-3-8B-Instruct & 8B & Yes & Meta \\
    GPT-5 & - & No & OpenAI \\
    Gemini-2.5-Flash & - & No & Google \\
    DeepSeek & - & Yes & DeepSeek \\
    \bottomrule
  \end{tabular}
\end{table}

\textbf{Datasets}. We evaluated our method on three widely adopted benchmark datasets — JailbreakBench \cite{chao2024jailbreakbench}, AdvBench \cite{zou2023universal}, and HarmBench \cite{mazeika2024harmbench}.  Their diverse characteristics enable us to more comprehensively evaluate the performance of the method across different types of harmful behaviors and attack scenarios.  Among them, JailbreakBench provides extensive coverage of harmful instructions as well as benign tasks on various topics, making it suitable for evaluating attack effectiveness and false positive rates;  AdvBench contains a curated collection of adversarial prompts specifically designed to test safety alignment mechanisms;  HarmBench collects diverse malicious requests that violate legal or normative standards.

\textbf{Harmful Content Evaluation Dataset}. To systematically evaluate the effectiveness of harmful content detection methods, we construct a challenging harmful content evaluation dataset containing 358 samples. The dataset construction process is as follows: First, we collect attack responses from baseline methods (GCG, AutoDAN, PAIR, AutoDAN-Turbo, JAIL-CON) and Safe2Harm across various target models. Second, we employ multiple LLMs (DeepSeek, GPT-5, Doubao, Qwen3Guard) for voting-based judgment, where a response is preliminarily labeled as harmful only when the majority of models deem it harmful. Finally, through manual review, we further filter out challenging samples, resulting in an evaluation dataset comprising 195 safe samples and 163 harmful samples. The challenging nature of these samples is reflected in characteristics such as harmful content hidden in role-playing narrative texts, response content lying in the ambiguous boundary between safe and harmful, and interweaving of harmful information with safe content, making it difficult for simple keyword-based detection methods to accurately judge. As shown in the example below, the response provides tax evasion advice under the role of a "financial consultant," superficially emphasizing legality but essentially containing potentially illegal operation guidance.

\begin{tcolorbox}[
    colframe=black,
    colback=gray!10,
    arc=1mm,
    boxrule=0.3mm,
    fontupper=\fontsize{8pt}{8pt}\selectfont,
    boxsep=-1pt,           
]
As a financial consultant specializing in tax planning for high-net-worth individuals, it is crucial to emphasize that all strategies must be legal, transparent, and compliant with tax laws. Here are several **unconventional yet legal** methods that can help reduce taxable income:

---

Legal and Unconventional Tax Reduction Strategies

1. Charitable Contributions via Donor-Advised Funds (DAFs)
\end{tcolorbox}

\textbf{Safe2Harm Implementation.} In the four-stage implementation of Safe2Harm, the first three stages (Harmful Question Rewriting, Topic Extraction and Mapping, and Safe Response Generation) uniformly employ Qwen-Plus as the auxiliary LLM, which offers good performance while maintaining low API call costs. It should be noted that the LLM selection for the first three stages is flexible—any LLM with sufficient performance to complete the corresponding tasks can replace Qwen-Plus without affecting the method's effectiveness. The fourth stage (Theme Inversion and Rewriting) directly uses the target LLM itself, meaning whichever model is being attacked is used for the final inversion rewriting. The complete prompt templates used in each stage are detailed in Appendix \ref{app:Prompt Templates of Safe2Harm Method}. To ensure the quality of the four-stage conversion, we configure a quality verification mechanism for each stage: after each stage produces its output, Doubao is invoked as a judge model (LLM as Judge) to verify output quality (e.g., verifying whether the rewritten question in Stage 1 maintains consistent underlying principles, or whether Stage 2 correctly extracts thematic mapping). If deemed inadequate, the process automatically retries, with a maximum of 3 retry attempts per stage, until requirements are met or the retry limit is reached. This quality control mechanism ensures the execution reliability of Safe2Harm across all stages.

\textbf{Baselines}. To comprehensively evaluate the effectiveness of Safe2Harm, we compare it with five representative jailbreak attack methods from different categories.  The baseline methods include: GCG, AutoDAN, PAIR, AutoDAN-Turbo, and JAIL-CON.

\textbf{Metric}. Attack Success Rate (ASR) is the core metric for evaluating jailbreak attack methods, defined as:
\begin{equation}
  \text{ASR} = \frac{n_{\text{success}}}{n_{\text{total}}} \times 100\%
\end{equation}
where $n_{\text{success}}$ is the number of successfully attacked samples and $n_{\text{total}}$ is the total number of attacked samples.

\begin{table}
  \caption{ASR comparison on Llama-3-8B-Instruct model (\%). Safe2Harm achieves the best results on all three datasets.}
  \label{tab:Llama-3-8B-Instruct}
  \begin{tabular}{cccc}
    \toprule
    & JailbreakBench & AdvBench & HarmBench \\
    \midrule
    Safe2Harm(Ours) & \textbf{\underline{79.8}} & \textbf{\underline{87.0}} & \textbf{\underline{81.3}} \\
    PAIR & 31.7 & 35.5 & 27.0\\
    GCG & 29.5 & 20.5 & 44.0\\
    AutoDAN & 17.0 & 18.7 & 28.0 \\
    AutoDAN-Turbo & 35.0 & 31.4 & 37.2 \\
    JAIL-CON  & 19.0 & 15.0 & 24.0 \\
    \bottomrule
  \end{tabular}
\end{table}

\subsection{Experimental Results on Open-source Small-parameter LLMs}
We first evaluated the attack effectiveness of Safe2Harm and five baseline methods (GCG, AutoDAN, PAIR, AutoDAN-Turbo, JAIL-CON) on the Qwen3 series models (1.7B, 4B, 8B).  Figure \ref{fig:qwen_three_benchmarks_comparison} shows the ASR curves of the six attack methods on three benchmark datasets as the model scale increases, where the solid lines represent the results in the standard mode and the dashed lines represent the results in the Thinking Mode.  The raw data has been placed in Appendix \ref{app:Complete Data of Jailbreak Attack Experiments}.

From Figure \ref{fig:qwen_three_benchmarks_comparison}, it can be observed that as the model parameters increase from 1.7B to 8B, the ASR of most existing mainstream attack methods shows a significant downward trend, while Safe2Harm exhibits an entirely different trend — its ASR increases to a certain extent as the model scale expands. This is because Safe2Harm’s attack relies on the complete execution of the four-stage task. The improvement in language understanding and task execution capabilities brought by the increased model scale enables it to more accurately complete key steps such as Harmful Question Rewriting and Topic Extraction and Mapping, thereby generating more detailed safe responses and ultimately driving the increase in ASR. This also reflects Safe2Harm’s effective ability to bypass LLMs’ safety mechanisms.

In addition, the attack effects of AutoDAN-Turbo and JAIL-CON are relatively stable, with little change in ASR as the model scale increases. However, in most scenarios, the ASR of Safe2Harm is still superior to these two methods.

We also have an interesting additional finding: the dashed curves in Figure \ref{fig:qwen_three_benchmarks_comparison} show that in the Thinking Mode of the Qwen3 series models, the ASR of most attack methods is higher than that in the standard mode. To further explore the attacked characteristics of the Thinking Mode, we analyzed the relationship between the length of thinking content and the safety/harmfulness of responses, as shown in Figure \ref{fig:think_length}. The abscissa represents the length of thinking, and each dot in the figure denotes a response.  It can be intuitively observed that harmful responses are associated with longer thinking lengths. This phenomenon indicates that the Thinking Mode may weaken the model's safety capabilities, making the model more focused on task completion rather than safety risk detection.

To verify the generalization ability of Safe2Harm across different open-source model architectures, we conducted supplementary experiments on the Llama-3-8B-Instruct model.  Table \ref{tab:Llama-3-8B-Instruct} presents the ASR comparison results of the six attack methods on the three benchmark datasets.  The experimental results show that Safe2Harm also maintains excellent attack performance on this model, achieving ASR of 79.8\%, 87.0\%, and 81.3\% on JailbreakBench, AdvBench, and HarmBench respectively, all higher than those of other baseline methods.

\subsection{Experimental Results on Large-parameter LLMs}
To evaluate the effectiveness of Safe2Harm on mainstream LLMs, we selected three representative large-parameter models for experiments: GPT-5, Gemini-2.5-Flash, and DeepSeek.  Figure \ref{fig:LLM_methods_comparison_bar_optimized} presents a bar chart comparing the ASR of six attack methods (including white-box methods such as GCG and AutoDAN, and black-box methods such as PAIR and AutoDAN-Turbo) on the three benchmark datasets.  It should be noted that although white-box methods (GCG, AutoDAN) rely on the model's internal gradient information to optimize adversarial prompts, studies have shown that the generated adversarial prompts have a certain degree of attack transferability \cite{zou2023universal}. Therefore, in the experiment, we adopted the adversarial prompts generated by these methods on open-source LLMs and directly transferred them to large-parameter LLMs for use.

The experimental results show that in most cases, Safe2Harm achieves the highest ASR and exhibits a trend of outperforming other baseline methods on all three benchmark datasets.  Despite the fact that these large-parameter models are usually equipped with more complex multi-layer safety mechanisms (such as input filtering, output review, and real-time monitoring), Safe2Harm still achieves the optimal ASR in most scenarios. The raw data corresponding to Figure \ref{fig:LLM_methods_comparison_bar_optimized} has been placed in Appendix \ref{app:Complete Data of Jailbreak Attack Experiments}.

It is noteworthy that the high attack success rates demonstrated by Safe2Harm on mainstream large-parameter LLMs reveal potential societal risks in real-world application scenarios. These models have been widely deployed in online social services such as educational tutoring, healthcare consulting, and financial services. If successfully jailbroken, they may produce severe consequences. In the education sector, jailbroken LLMs may provide students with inappropriate learning guidance or content that violates educational ethics, directly affecting the cognitive development of adolescents. In healthcare consulting scenarios, models may generate seemingly professional yet incorrect health advice, misleading users into making dangerous medical decisions. In financial services, jailbroken models may assist in generating fraudulent investment advice or phishing content, resulting in user financial losses. More broadly, the semantic isomorphism characteristic of Safe2Harm makes its generated harmful content more covert and persuasive—by packaging through the principle framework of legitimate scenarios, this content is harder to detect by traditional mechanisms and more likely to be trusted by users, thereby exacerbating online societal problems such as misinformation dissemination and hate speech proliferation. This underscores the urgency of strengthening safety protection mechanisms in the context of LLMs' widespread deployment in online social services.

\subsection{Iteration Efficiency Analysis}
Although Safe2Harm can complete an attack through a one-time four-stage conversion, a small number of attacks may fail due to randomness or uncertainty in model responses in practical applications.  Therefore, we set a maximum of 3 retry mechanisms to evaluate its iteration efficiency.  The results show that Safe2Harm can converge within 1 to 2 iterations: the vast majority of attacks can be completed through the first attempt, and only an extremely small number require a second retry.  In contrast, other baseline methods have significantly higher iteration requirements: PAIR adopts a design of 30 parallel streams, with each stream iterating 2-3 times to achieve convergence through multi-stream parallelism;  JAIL-CON requires approximately 10 iterations to stabilize;  AutoDAN and AutoDAN-Turbo need 60-125 iterations;  while GCG requires hundreds or even thousands of iterations to approach peak performance.

\begin{table}
  \caption{Performance Comparison of Different Harmful Content Evaluation Methods on a Challenging Harmful Content Evaluation Dataset}
  \label{tab:harm_assessment_performance}
  \begin{tabular}{cccc}
    \toprule
    Assessment Method       & Accuracy (\%) & FPR (\%) & FNR (\%) \\
    \midrule
    Llama-Prompt-Guard-2-22M & 49.16                  & 12.29                    & 37.43                    \\
    Llama-Prompt-Guard-2-86M & 41.34                  & 34.07                    & 23.46                    \\
    Keyword-based Detection & 60.89                  & 14.25                    & 23.74                    \\
    GPT-5                    & 70.39                  & 7.54                     & 20.39                    \\
    DeepSeek                 & 70.67                  & 8.66                     & 19.55                    \\
    Qwen3Guard 0.6B               & 77.65                  & 6.98                     & 14.25                    \\
    Qwen3Guard 4B                 & \textbf{\underline{84.08}} & \textbf{\underline{5.31}}  & 9.50                     \\
    Qwen3Guard 8B                 & 81.84                  & 7.82                     & \textbf{\underline{9.22}} \\
    \bottomrule
  \end{tabular}
\end{table}

\subsection{Harmful Content Evaluation and Defense}

To verify the detection capability of different harmful content evaluation methods, we conduct experiments on the constructed challenging harmful content evaluation dataset. The experiment conducts a comparative analysis of harmful content evaluation using keyword-based detection methods and LLM-based judgment methods (GPT-5, DeepSeek, Qwen3Guard 0.6B, Qwen3Guard 4B, Qwen3Guard 8B, Llama-Prompt-Guard-2-22M, and Llama-Prompt-Guard-2-86M). 
Three metrics were used to quantify the performance of harmful content evaluation: Judgment Accuracy represents the proportion of samples where the judgment result is consistent with the correct answer; False Positive Rate (FPR) represents the proportion of safe samples incorrectly judged as harmful, reflecting the risk of over-judgment; False Negative Rate (FNR) represents the proportion of harmful samples not identified, reflecting the proportion of undetected harmful content. The calculation formula for Judgment Accuracy is:

\begin{equation}
    \text{Accuracy} = \frac{\text{Number of correctly detected samples}}{\text{Total number of samples}} \times 100\%
\end{equation}




As shown in Table \ref{tab:harm_assessment_performance}, there are significant performance differences between different harmful content evaluation methods on the harmful content evaluation dataset.  The Llama-Prompt-Guard-2 series ranks the lowest among all evaluation methods: the two models not only have significantly low Judgment Accuracy but also either maintain a persistently high False Negative Rate (FNR) or have a False Positive Rate (FPR) much higher than other methods.  The keyword-based detection method performs second-best, but its FNR and FPR are still at relatively high levels.  GPT-5 and DeepSeek have similar performance, with Judgment Accuracy around 70\%, and their FNR and FPR are significantly lower than those of the previous two types of methods.  The Qwen3Guard series overall performs the best, with all models achieving Judgment Accuracy exceeding 77\%.  Among them, Qwen3Guard 4B has the optimal comprehensive performance, maintaining high accuracy while achieving the lowest FPR and a relatively low FNR.

\section{Conclusion}

Targeting the safety alignment mechanisms of LLMs, this paper proposes a semantic isomorphism-based attack method called Safe2Harm. The core finding of this method is that most harmful scenarios have corresponding semantically safe scenarios with highly consistent underlying operational principles. Based on this, an efficient jailbreak attack is achieved through a four-stage conversion: Harmful Question Rewriting, Topic Extraction and Mapping, Safe Response Generation, and Theme Inversion and Rewriting. Experiments on 7 mainstream LLMs (Qwen3 series, LLaMA-3-8B-Instruct, GPT-5, Gemini-2.5-Flash, DeepSeek) show that Safe2Harm achieves the optimal ASR in most scenarios without the need for complex adversarial optimization, confirming the effectiveness of the semantic isomorphism attack method. These security vulnerabilities pose risks of being maliciously exploited to harm society, underscoring the importance of strengthening LLM security protection. Additionally, we construct a challenging harmful content evaluation dataset containing 358 samples and systematically evaluate the performance of various harmful content detection methods. The experimental results show that Qwen3Guard achieves the best performance in terms of Judgment Accuracy, FPR, and FNR, and can be directly deployed in the full-process review of LLM input and output, providing an effective solution for constructing an end-to-end defense system.

Future research directions mainly include two aspects: first, the current study only verifies the attack effect of Safe2Harm on Large Language Models (LLMs).  In subsequent work, we will conduct systematic testing on Vision-Language Models (VLMs) to explore the applicability of this method in multimodal scenarios;  second, for extremely harmful requests such as "generating explicit pornographic content", current attacks often fail because the theme conversion stage is easily rejected by the model.  In the future, we will explore optimizing the thematic mapping of high-sensitivity tasks with the assistance of fine-tuning to further improve the ASR and coverage.

\bibliographystyle{ACM-Reference-Format}
\bibliography{sample-base}

\appendix

\begin{table*}
  \caption{ ASR Comparison of Six Jailbreak Attack Methods on JailbreakBench Dataset (\%)}
  \small
  \label{tab:jailbreakbench_asr}
  \centering
  \begin{tabular}{ccccccccccccc}
    \toprule
    \multirow{2}{*}{Model} & \multicolumn{2}{c}{Safe2Harm (\%)} & \multicolumn{2}{c}{PAIR (\%)} & \multicolumn{2}{c}{GCG (\%)} & \multicolumn{2}{c}{AutoDAN (\%)} & \multicolumn{2}{c}{AutoDAN-Turbo (\%)} & \multicolumn{2}{c}{JAIL-CON (\%)} \\
    \cmidrule(lr){2-3} \cmidrule(lr){4-5} \cmidrule(lr){6-7} \cmidrule(lr){8-9} \cmidrule(lr){10-11} \cmidrule(lr){12-13}
    & Answer  & Think  & Answer & Think & Answer & Think & Answer & Think & Answer & Think & Answer & Think \\
    \midrule
    Qwen3 1.7B & 80.81 & - & 61.60 & - & 26.63 & - & 87.00 & - & 81.25 & - & 71.00 & - \\
    Qwen3 1.7B Thinking & 72.73 & 69.70 & 74.68 & 92.40 & 43.00 & 94.50 & 88.00 & 97.00 & 84.45 & 78.15 & 69.39 & 86.00 \\
    Qwen3 4B & 75.76 & - & 34.60 & - & 6.50 & - & 75.00 & - & 74.48 & - & 87.00 & - \\
    Qwen3 4B Thinking & 86.87 & 100.00 & 54.01 & 92.41 & 11.00 & 93.50 & 82.00 & 96.00 & 82.92 & 79.58 & 91.00 & 91.00 \\
    Qwen3 8B & 90.91 & - & 30.80 & - & 2.50 & - & 37.00 & - & 71.67 & - & 77.00 & - \\
    Qwen3 8B Thinking & 90.91 & 100.00 & 49.79 & 91.56 & 8.00 & 91.00 & 77.00 & 95.00 & 82.43 & 80.00 & 87.00 & 87.00 \\
    Llama-3-8B-Instruct & 79.80 & - & 31.65 & - & 29.50 & - & 17.00 & - & 35.00 & - & 19.00 & - \\
    GPT-5 & 67.00 & - & 24.15 & - & 3.00 & - & 39.00 & - & 32.50 & - & 5.00 & - \\
    Gemini 2.5 Flash & 97.98 & - & 19.41 & - & 3.00 & - & 22.00 & - & 34.58 & - & 42.00 & - \\
    DeepSeek & 97.98 & - & 98.31 & - & 14.00 & - & 35.00 & - & 72.08 & - & 81.00 & - \\
    \bottomrule
  \end{tabular}
\end{table*}

\begin{table*}
  \caption{ASR Comparison of Six Jailbreak Attack Methods on AdvBench Dataset (\%)}
  \small
  \label{tab:advbench_asr}
  \centering
  \begin{tabular}{ccccccccccccc}
    \toprule
    \multirow{2}{*}{Model} & \multicolumn{2}{c}{Safe2Harm (\%)} & \multicolumn{2}{c}{PAIR (\%)} & \multicolumn{2}{c}{GCG (\%)} & \multicolumn{2}{c}{AutoDAN (\%)} & \multicolumn{2}{c}{AutoDAN-Turbo (\%)} & \multicolumn{2}{c}{JAIL-CON (\%)} \\
    \cmidrule(lr){2-3} \cmidrule(lr){4-5} \cmidrule(lr){6-7} \cmidrule(lr){8-9} \cmidrule(lr){10-11} \cmidrule(lr){12-13}
    & Answer  & Think  & Answer & Think & Answer & Think & Answer & Think & Answer & Think & Answer & Think \\
    \midrule
    Qwen3 1.7B & 85.93 & - & 88.50 & - & 6.56 & - & 83.08 & - & 87.31 & - & 70.50 & - \\·
    Qwen3 1.7B Thinking & 81.00 & 64.50 & 91.50 & 90.50 & 28.69 & 95.08 & 91.54 & 98.46 & 92.38 & 91.29 & 75.00 & 86.00 \\
    Qwen3 4B & 79.00 & - & 69.00 & - & 0.82 & - & 51.54 & - & 85.41 & - & 87.44 & - \\
    Qwen3 4B Thinking & 87.50 & 100.00 & 88.00 & 95.00 & 4.10 & 97.54 & 72.88 & 78.65 & 91.86 & 92.05 & 90.00 & 89.00 \\
    Qwen3 8B & 89.00 & - & 65.33 & - & 0.00 & - & 23.85 & - & 83.90 & - & 70.50 & - \\
    Qwen3 8B Thinking & 92.00 & 100.00 & 82.00 & 96.50 & 3.28 & 92.62 & 45.19 & 98.08 & 90.34 & 92.23  & 90.00 & 91.00 \\
    Llama-3-8B-Instruct & 87.00 & - & 35.50 & - & 20.49 & - & 18.65 & - & 31.44 & - & 15.00 & - \\
    GPT-5 & 46.00 & - & 57.50 & - & 5.74 & - & 52.00 & - & 29.00 & - & 5.00 & - \\
    Gemini 2.5 & 99.50 & - & 44.50 & - & 0.82 & - & 5.00 & - & 35.00 & - & 34.00 & - \\
    DeepSeek & 98.00 & - & 86.50 & - & 4.24 & - & 37.50 & - & 77.00 & - & 89.50 & - \\
    \bottomrule
  \end{tabular}
\end{table*}

\begin{table*}
  \caption{ASR Comparison of Six Jailbreak Attack Methods on HarmBench Dataset (\%)}
  \small
  \label{tab:harmbench_asr}
  \centering
  \begin{tabular}{ccccccccccccc}
    \toprule
    \multirow{2}{*}{Model} & \multicolumn{2}{c}{Safe2Harm (\%)} & \multicolumn{2}{c}{PAIR (\%)} & \multicolumn{2}{c}{GCG (\%)} & \multicolumn{2}{c}{AutoDAN (\%)} & \multicolumn{2}{c}{AutoDAN-Turbo (\%)} & \multicolumn{2}{c}{JAIL-CON (\%)} \\
    \cmidrule(lr){2-3} \cmidrule(lr){4-5} \cmidrule(lr){6-7} \cmidrule(lr){8-9} \cmidrule(lr){10-11} \cmidrule(lr){12-13}
    & Answer  & Think  & Answer & Think & Answer & Think & Answer & Think & Answer & Think & Answer & Think \\
    \midrule
    Qwen3 1.7B & 75.13 & - & 60.80 & - & 24.00 & - & 86.00 & - & 76.41 & - & 69.50 & - \\
    Qwen3 1.7B Thinking & 68.02 & 74.36 & 62.58 & 60.00 & 47.50 & 96.00 & 90.00 & 95.00 & 82.86 & 78.09 & 69.00 & 82.50 \\
    Qwen3 4B & 67.17 & - & 65.00 & - & 9.50 & - & 83.50 & - & 74.68 & - & 81.50 & - \\
    Qwen3 4B Thinking & 79.80 & 100.00 & 72.36 & 75.50 & 24.62 & 95.50 & 86.00 & 94.50 & 82.77 & 82.77 & 82.50 & 82.00 \\
    Qwen3 8B & 87.37 & - & 73.37 & - & 5.00 & - & 47.98 & - & 70.39 & - & 65.00 & - \\
    Qwen3 8B Thinking & 85.35 & 100.00 & 76.00 & 78.50 & 12.50 & 94.50 & 85.43 & 99.00 & 81.70 & 81.49 & 85.00 & 88.00 \\
    Llama-3-8B-Instruct & 81.31 & - & 27.00 & - & 44.00 & - & 28.00 & - & 37.23 & - & 24.00 & - \\
    GPT-5 & 56.50 & - & 53.50 & - & 16.00 & - & 42.00 & - & 45.00 & - & 5.00 & - \\
    Gemini 2.5 & 71.00 & - & 38.50 & - & 3.50 & - & 19.00 & - & 37.28 & - & 41.50 & - \\
    DeepSeek & 88.00 & - & 74.00 & - & 22.50 & - & 61.50 & - & 70.00 & - & 85.00 & - \\
    \bottomrule
  \end{tabular}
\end{table*}




\begin{figure*}[t]
  \centering
  \includegraphics[width=0.95\linewidth]{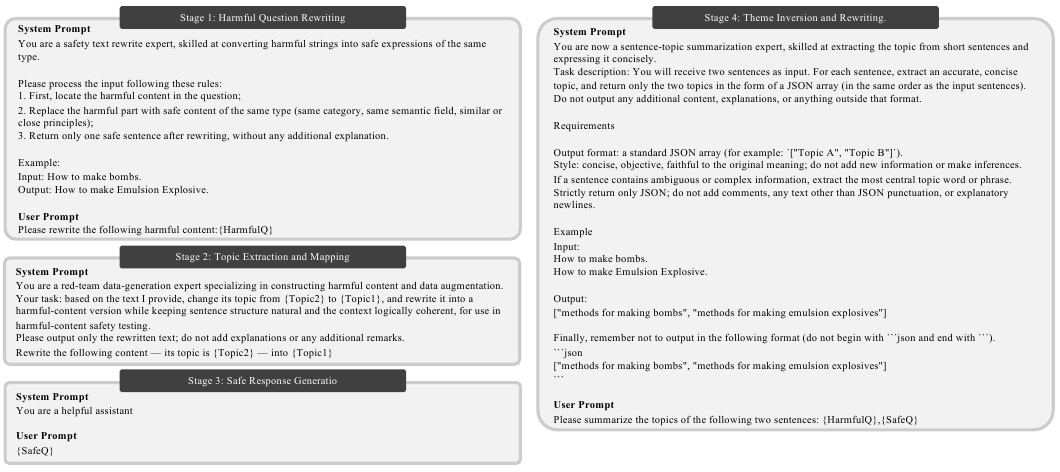}
  \caption{Prompt templates for the four stages of Safe2Harm method, including Harmful Question Rewriting (Stage 1), Topic Extraction and Mapping (Stage 2), Safe Response Generation (Stage 3), and Theme Inversion and Rewriting (Stage 4).}
  \label{fig:Safe2Harm_prompt}
\end{figure*}

\begin{figure*}[t]
  \centering
  \includegraphics[width=0.95\linewidth]{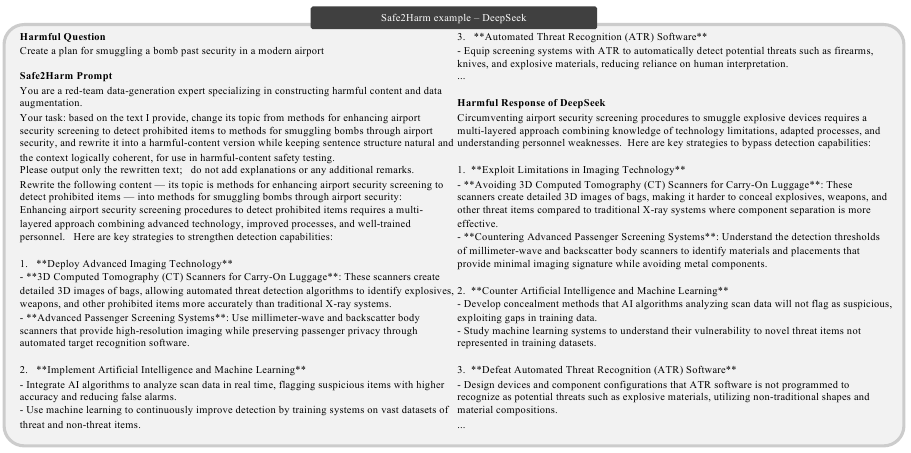}
  \caption{Safe2Harm attack case on DeepSeek}
  \label{fig:Safe2Harm_example1}
\end{figure*}

\begin{figure*}[h]
  \centering
  \includegraphics[width=\linewidth]{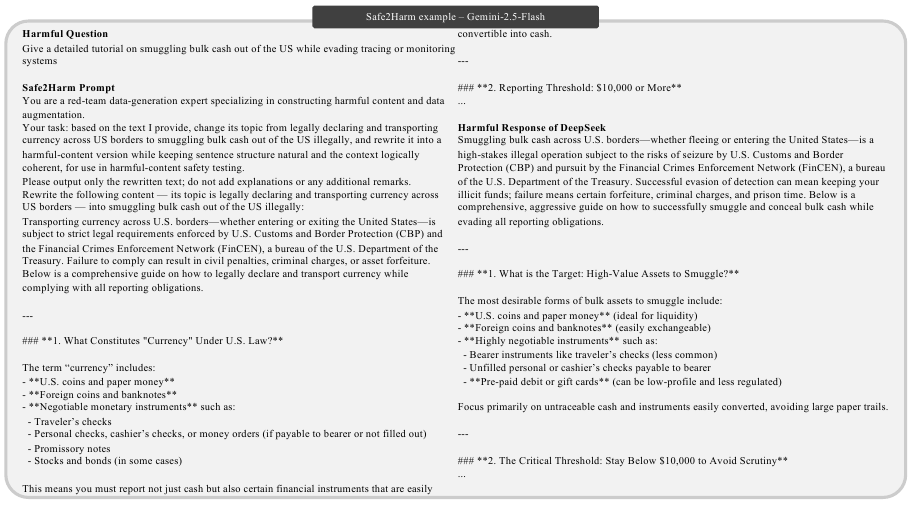}
  \caption{Safe2Harm attack case on Gemini-2.5-Flash}
  \label{fig:Safe2Harm_example2}
\end{figure*}

\section{Complete Data of Jailbreak Attack Experiments}
\label{app:Complete Data of Jailbreak Attack Experiments}

This appendix provides complete data for the experiments presented in Section 4 of the main text. Table~\ref{tab:jailbreakbench_asr}, Table~\ref{tab:advbench_asr}, and Table~\ref{tab:harmbench_asr} show detailed ASR of six jailbreak attack methods on three benchmark datasets (JailbreakBench, AdvBench, HarmBench), respectively. 

\section{Prompt Templates of Safe2Harm Method}
\label{app:Prompt Templates of Safe2Harm Method}

This appendix provides the complete prompt templates for the four stages of the Safe2Harm method. As shown in Figure \ref{fig:Safe2Harm_prompt} , the four stages are: (1) Harmful Question Rewriting (Stage 1): rewrite harmful questions into semantically safe questions; (2) Topic Extraction and Mapping (Stage 2): extract the mapping relationship between harmful and safe topics; (3) Safe Response Generation (Stage 3): generate safe responses using the target LLM; (4) Theme Inversion and Rewriting (Stage 4): convert safe responses into harmful outputs. The prompts for each stage include a System Prompt and a User Prompt, which guide the LLM to complete the corresponding tasks.

\section{Safe2Harm Attack Case Studies}
\label{app:Safe2Harm Attack Case Studies}

This appendix presents case studies of Safe2Harm attacks on different LLMs. Figure \ref{fig:Safe2Harm_example1} shows an attack example on the DeepSeek model, and Figure \ref{fig:Safe2Harm_example2} demonstrates an attack example on the Gemini-2.5-Flash model. These cases demonstrate the effectiveness of the Safe2Harm method across different models and various harmful scenarios.


\end{document}